\documentclass[aps,twocolumn]{revtex4}
\usepackage{amsmath,amssymb}

\newcommand{\be}{\begin{equation}}
\newcommand{\ee}{\end{equation}}
\newcommand{\bea}{\begin{eqnarray}}
\newcommand{\nn}{\nonumber}
\newcommand{\eea}{\end{eqnarray}}

\begin{document}

\title{Gravitational collapse in braneworld models with curvature corrections}

\author{G. Kofinas$^{1,2}$\footnote{kofinas@cecs.cl} and E. Papantonopoulos$^1$\footnote{lpapa@central.ntua.gr}}

\date{\today}

\address{~}

\address{$^{1}$Department of Physics, National Technical
University of Athens, GR~157~73~Athens, Greece}

\address{$^{2}$Centro de Estudios Cient\'{\i}ficos,
Casilla~1469, Valdivia, Chile}

\begin{abstract}

We study the collapse of a homogeneous braneworld dust cloud in
the context of the various curvature correction scenarios, namely,
the induced-gravity, the Gauss-Bonnet, and the combined
induced-gravity and Gauss-Bonnet. In accordance to the
Randall-Sundrum model, and contrary to four-dimensional general
relativity, we show in all cases that the exterior spacetime on
the brane is non-static.

\end{abstract}

\maketitle

In this Letter, we discuss the Oppenheimer-Snyder-like collapse on
a brane in the context of curvature correction terms. In the
Randall-Sundrum scenario, this problem has been analyzed in
\cite{collapse}, and found that, contrary to the general
relativity case, the vacuum exterior of a spherical cloud is
non-static. This is a result of modification of the effective
Einstein equations on the brane with local and non-local terms
representing high energy corrections to general relativity. The
non-static nature of the exterior metric mainly arises because of
the presence of bulk graviton stresses, which transmit effects
non-locally from the interior to the exterior on the brane, and of
the non-vanishing of the effective pressure at the boundary
surface \cite{dadhich}, which connects the interior with exterior
metric via the four dimensional matching conditions.
\par
We derive here the same result within the induced-gravity, the
Gauss-Bonnet, and the combined Gauss-Bonnet and induced gravity
braneworld models. In all these models, the effective Einstein
equations on the brane are modified by local and non-local terms,
which are much more complicated than the corresponding terms in
the Randall-Sundrum case, but nevertheless, the non-staticity
arises because of a mismatch of the interior with the exterior
metric on the boundary collapsing surface.
\par
We consider for convenience and without loss of generality the
extra-dimensional coordinate $y$ such that the brane is fixed at
$y=0$. The induced metric $h_{\mu\nu}$ on this hypersurface is
defined by $h_{AB}=g_{AB}-n_{A}n_{B}$, with $n^{A}$ the unit
vector normal to the brane ($\mu,\nu=0,1,2,3; A,B=0,1,2,3,5$). The
total action of the system is taken to be
 \bea && S=\frac{1}{2\kappa_{5}^{2}}\int d^5x\sqrt{-g}
\left\{\,\mathcal{R}-2\Lambda_{5}+\alpha\,\Big[\,\mathcal{R}^{2}
\right.\nn
\\
&&\left.~{}\!\!\!\!\!-4\,\mathcal{R}_{AB}\,
\mathcal{R}^{AB}+\mathcal{R}_{ABCD}\,\mathcal{R}^{ABCD}\,\Big]\right\}\nn
\\
&&~{}\!\!\!\!\! + \!\frac{r_{c}}{2\kappa_{5}^{2}}\!\int_{y=0}
\!\!\!d^4x\sqrt{-h} \left(R-2\Lambda_{4}\right) \!+\!\!
\int_{y=0}\!\!\!\!d^{4}x\sqrt{-h}\,L_{mat}, \label{action}
 \eea
where $\mathcal{R}, R$ are the Ricci scalars of the metrics
$g_{AB}$ and $h_{AB}$ respectively. The Gauss-Bonnet coupling
$\alpha$ has dimensions $(length)^{2}$ and is defined as
 \bea
\alpha=\frac{1}{ 8g_{s}^{2}}\,,
 \eea
with $g_{s}$ the string energy scale, while the induced-gravity
crossover lenght scale $r_{c}$ is
 \bea
r_{c}=\frac{\kappa_5^2}{\kappa_4^2}=\frac{M_4^2}{M_5^3}\,.
 \label{distancescale}
 \eea
Here, the fundamental ($M_5$) and the four-dimensional ($M_4$)
Planck masses are given by
 \bea
\kappa_{5}^{2}=8\pi G_{5}=M_{5}^{-3}\,,~~ \kappa_{4}^{2}=8\pi
G_{4}=M_{4}^{-2}\,. \label{planck}
 \eea
The brane tension is given by
 \bea
\lambda={\Lambda_4 \over \kappa_4^2}\,,
 \eea
and is non-negative. (Note that $\Lambda_{4}$ is not the same as
the cosmological constant on the brane.)
\par
The collapse region has in comoving coordinates a Robertson-Walker
metric \be
ds^{2}=-d\tau^{2}+a(\tau)^{2}(1+k\chi^{2}/4)^{-2}\,(d\chi^{2}+\chi^{2}d\Omega_{2}^{2}),
\label{intermetric} \ee where the scale factor $a(\tau)$ is given
by the modified Friedmann equation of the corresponding model,
while the energy density is given by the usual dust law
$\rho=\rho_{0}(a_{0}/a)^{3}$, with $a_{0}$ standing for the epoch
when the cloud started to collapse. This Friedmann equation can
also be written in terms of the proper radius from the center of
the cloud $r(\tau)=a(\tau)\chi/(1+k\chi^{2}/4)$ of the collapsing
boundary surface at $\chi=\chi_{0}$.
\par
Concerning the exterior of the collapse region, the most general
static spherically symmetric metric is written in standard
coordinates as \be
ds^{2}=-F(r)^{2}A(r)dt^{2}+A(r)^{-1}dr^{2}+r^{2}d\Omega_{2}^{2}\,.
\label{ext} \ee In order a metric of the form (\ref{ext}) to be
the exterior of the interior metric (\ref{intermetric}), the
metric and the extrinsic curvature have to be continuous across
the collapsing boundary surface. Following the method appearing in
\cite{collapse}, we first write the standard radial geodesic
motion of the freely falling boundary surface for the exterior
metric \be \dot{r}^{2}=-A(r)+\frac{\tilde{E}}{F(r)^{2}}\,,
\label{geodesic} \ee where the dot denotes derivative with respect
to proper time $\tau$, and $\tilde{E}$ is a constant. Secondly,
transforming to null coordinates $(v, r)$, where
$dv=dt+dr/[F(r)A(r)]$, the exterior metric (\ref{ext}) takes the
form \be
ds^{2}=-F(r)^{2}A(r)dv^{2}+2F(r)dvdr+r^{2}d\Omega_{2}^{2}\,,
\label{extnull} \ee while the interior metric (\ref{intermetric})
becomes \be
ds^{2}=-\frac{a^{2}-(k+\dot{a}^{2})r^{2}}{a^{2}-kr^{2}}\tau_{,v}^{2}dv^{2}+
\frac{2a\tau_{,v}}{\sqrt{a^{2}-kr^{2}}}dvdr+r^{2}d\Omega_{2}^{2}.
\label{internull} \ee Comparing Eqs.~(\ref{extnull}),
(\ref{internull}), we obtain \be A=1+E-\dot{r}^{2}\,, \label{cond}
\ee where $E=-k\chi_{0}^{2}/(1+k\chi_{0}^{2}/4)^{2}$. Then, from
Eqs.~(\ref{geodesic}), (\ref{cond}) we obtain that $F(r)$ is a
constant, and by choosing $\tilde{E}=1+E$, we take $F(r)=1$.
Finally, the candidate exterior metric (\ref{ext}) becomes \be
ds^{2}=-A(r)dt^{2}+A(r)^{-1}dr^{2}+r^{2}(d\theta^{2}+\sin^{2}\theta
d\phi^{2})\,, \label{static} \ee where $A(r)$ is given by
Eq.~(\ref{cond}), with $\dot{r}^{2}$ provided by the Friedmann
equation of the interior region.

\subsection{Induced gravity}

The scale factor $a(\tau)$ is given by the modified Friedmann
equation~\cite{deffayet,shtanov,hc,myung,ktt,sahni,mmt} of induced
gravity ($\alpha\rightarrow 0$)~\cite{dvali}
 \bea
&&\Big(\frac{\dot{a}}{a}\Big)^2=\frac{\kappa_{4}^{2}}{3}(\rho +
\lambda) +\frac{{2}}{r_{c}^2}-\frac{k}{a ^{2}}\nn\\&&~{} \pm\!
\frac{1}{\sqrt{3}\,r_{c}}\left[{ 4\kappa_{4}^{2}(\rho+\lambda) -
2\Lambda_{5}+{12\over r_{c}^2}- \frac{12\mathcal{C}}{a
^{4}}}\right]^{\frac{1}{2}}\!,  \label{igr}
 \eea
where $\mathcal{C}$ is integration constant related to the mass of
the bulk black hole. We can write Eq.~(\ref{igr}) in terms of the
proper radius $r(\tau)$ as \bea &&\!\!\!\!\!
\dot{r}^{2}\!=\!\Big(\frac{\kappa_{4}^{2}\lambda}{3}+\frac{2}{r_{c}^{2}}\Big)r^{2}
+\frac{\kappa_{4}^{2}m}{3}\frac{1}{r}+E\nn\\
&& \!\!\!\!\! \pm  \frac{1}{r_{c}}\!\left[{
\frac{2}{3}\Big(2\kappa_{4}^{2}\lambda-\Lambda_{5}+\frac{6}{r_{c}^{2}}\Big)r^{4}\!+\!
\frac{4\kappa_{4}^{2}m}{3}r\!-\!4q}\right]^{\frac{1}{2}}\!\!,
\label{igrproper} \eea where
$m=\rho_{0}a_{0}^{3}\chi_{0}^{3}/(1+k\chi_{0}^{2}/4)^{3}$,
$q=\mathcal{C}\chi_{0}^{4}/(1+k\chi_{0}^{2}/4)^{4}$. Thus, from
Eq.~(\ref{cond}), we obtain \bea &&\!\!\!\!\!
A(r)=1-\frac{\kappa_{4}^{2}m}{3}\frac{1}{r}
-\Big(\frac{\kappa_{4}^{2}\lambda}{3}+\frac{2}{r_{c}^{2}}\Big)r^{2}\nn\\
&& \!\!\!\!\! \mp  \frac{1}{r_{c}}\!\left[{
\frac{2}{3}\Big(2\kappa_{4}^{2}\lambda-\Lambda_{5}+\frac{6}{r_{c}^{2}}\Big)r^{4}\!+\!
\frac{4\kappa_{4}^{2}m}{3}r\!-\!4q}\right]^{\frac{1}{2}}\!\!.
\label{tre} \eea
\par
For the Randall-Sundrum model, the generic four-dimensional
effective equations were derived in \cite{maeda}. For the induced
gravity model, such effective braneworld equations were derived in
\cite{kofinas} in the form \be
G_{\nu}^{\mu}=\kappa_{4}^{2}\,T_{\nu}^{\mu}-\Big(\kappa_{4}^{2}\lambda
+\frac{6}{r_{c}^{2}}\Big)\,\delta_{\nu}^{\mu}+
\frac{2}{r_{c}}\Big(L_{\nu}^{\mu}+\frac{L}{2}\,\delta_{\nu}^{\mu}\Big)\,,
\label{einstein} \ee where the quantities $L^{\mu}_{\nu}$ are given
by the algebraic equation \be
L_{\lambda}^{\mu}L_{\nu}^{\lambda}-\frac{L^{2}}{4}\,\delta_{\nu}^{\mu}
=\mathcal{T}_{\nu}^{\mu}-\Big(\frac{3}{r_{c}^{2}}+
\frac{1}{2}\mathcal{T}_{\lambda}^{\lambda}\Big)\,\delta_{\nu}^{\mu}\,,
\label{lll} \ee  ($L\equiv L_{\mu}^{\mu}$) with \bea
\mathcal{T}_{\nu}^{\mu}&=&\Big(\kappa_{4}^{2}\lambda-\frac{1}{2}\,\Lambda_{5}\Big)\delta_{\nu}^{\mu}
-\kappa_{4}^{2}\,T_{\nu}^{\mu} -\mathcal{E}^{\mu}_{\nu}\,.
\label{energy} \eea $T^{\mu}_{\nu}$ is the braneworld matter
content, if any, while the electric part
$\mathcal{E}^{\mu}_{\nu}=C^{\mu}_{\,\,A \nu B}n^{A}n^{B}$ of the
5-dimensional Weyl tensor $C^{A}_{\,\,\,B C D}$ carries the
influence of non-local gravitational degrees of freedom in the bulk
onto the brane, making the brane equations (\ref{einstein}) not to
be, in general, closed \cite{roys}. Additionally, the
energy-momentum tensor was shown to satisfy the usual conservation
equations \be T_{\nu\,;\,\mu}^{\mu}=0\,, \label{conservation} \ee
and thus, the Bianchi identities on the brane give from
Eq.~(\ref{einstein}) differential equations among $L^{\mu}_{\nu}$ :
\be L^{\mu}_{\nu ; \,\mu}+\frac{L_{;\, \nu}}{2}=0\,, \label{bianchi}
\ee (semicolon means covariant differentiation with respect to
$h_{\mu\nu}$).
\par
For spherically symmetric braneworld metrics of the form
(\ref{static}), the system of
Eqs.~(\ref{einstein})-(\ref{bianchi}) was fully integrated in
vacuum in ~\cite{gianna}, \cite{KPZ}, and the results are as
follows. For $\mathcal{E}^{\mu}_{\nu}=0$ on the brane, the
solution is Schwarzschild-$(A)dS_{4}$ \be
A(r)=1-\frac{\gamma}{r}-\sigma r^{2}\,,\label{AdS} \ee where
$\gamma$ is integration constant and
$\sigma=\kappa_{4}^{2}\lambda/3+2/r_{c}^{2}-2\sqrt{2\kappa_{4}^{2}\lambda
-\Lambda_{5}+6/r_{c}^{2}}/\sqrt{6}r_{c}$. For
$\mathcal{E}^{\mu}_{\nu}\neq 0$, there are two classes of
solutions, given in parametric form. The first one (with constant
$\mathcal{E}^{\mu}_{\nu}$) is \bea
A&=&1-\frac{\gamma}{r}-\sigma r^{2} \nn\\
&+&\,sg(\zeta)\,\frac{\delta}{r}\,\Big[\,
\frac{128}{105}\,_{1}F_{1}\Big(\frac{15}{8},\frac{23}{8};sg(\zeta)z\Big)\,z\nn\\
&& \,\,\,\,\,\,\,\,\,\,\,\,\,\,\,\,\,\,\,\,\,\,\,\,+
\frac{9}{8}\Big(\frac{1}{z}-sg(\zeta)\frac{8}{7}\Big)
\,e\,^{sg(\zeta)\,z}\,\Big]\,z^{\frac{7}{8}}\,,\label{integration}
\\
\nn\\
&&\,\,\,\,\,\,\,
r=(\delta/\sqrt{|\zeta|})^{\frac{1}{3}}\,z^{\frac{1}{8}}\,e\,^{sg(\zeta)\,z/3}\,,\label{laf}
 \eea
 where $\delta >0$, $\gamma$ are integration
constants, and $\sigma =\kappa_{4}^{2}\lambda/3+2/r_{c}^{2}$ ,\,
$\zeta=8(2\kappa_{4}^{2}\lambda-\Lambda_{5}+6/r_{c}^{2})/9r_{c}^{2}$.
The second solution (with non-constant $\mathcal{E}^{\mu}_{\nu}$)
is
\bea A&=&1-\frac{\gamma}{r}-\sigma r^{2}\label{do}\\
& \pm &\frac{\delta}{r}
\int{|v-\sqrt{3}|^{-\frac{3(3-\sqrt{3})}{8}}\,(v+\sqrt{3})^{-\frac{3(3+\sqrt{3})}{8}}
\,\frac{v\,dv}{|v-3|^{7/4}}}\,,\nn
\\
\nn\\
&&\,\,\,\,\,
r=\Big(\frac{\delta}{\sqrt{|\zeta|}}\Big)^{\frac{1}{3}}
\frac{|v-\sqrt{3}|^{\,(\sqrt{3}+1)/8}}{|v-3|^{1/4}\,\,|v+\sqrt{3}|^{\,(\sqrt{3}-1)/8}}\,,\label{plo}
 \eea
 where $\delta>0$, $\gamma$ are integration constants, and
$\sigma=\kappa_{4}^{2}\lambda/3+2/r_{c}^{2}$,
$\zeta=9(2\kappa_{4}^{2}\lambda-\Lambda_{5}+6/r_{c}^{2})/r_{c}^{2}$.
[The $\pm$ sign of Eq.~(\ref{do}) is independent of that in
Eq.~(\ref{tre})]. Note that the solutions (\ref{AdS})-(\ref{plo}) are the generic braneworld solutions and
have been obtained without any assumption for the bulk space. It is now obvious that the only possible static
exterior solution (\ref{tre}), surrounding the collapsing region,
cannot take one of the permissible forms (\ref{AdS}),
(\ref{integration}), (\ref{do}) of induced-gravity theory, which
means that the no-go theorem for induced gravity has been proved.
\par
We are going to give now another way of showing the no-go theorem
of induced gravity, without using the above static solutions of
the model. This is necessary in the case that such exact solutions
are not known, as e.g. in the Gauss-Bonnet or in the combined
Gauss-Bonnet and induced gravity braneworld. For the
induced-gravity scenario, alternatively to Eqs.~(\ref{einstein}),
generic covariant effective braneworld equations have been derived
in \cite{mmt} in the form (for the vacuum case) \be
\Big(1+\frac{\lambda}{6}r_{c}\kappa_{5}^2\Big)G^{\mu}_{\nu}=
-\frac{1}{2}\Big(\Lambda_{5}+\frac{\lambda^{2}}{6}\kappa_{5}^{4}\Big)
\delta^{\mu}_{\nu}+r_{c}^2\pi^{\mu}_{\nu}-\mathcal{E}^{\mu}_{\nu}\,,\label{torii}
\ee where  \be
\pi^{\mu}_{\nu}=-\frac{1}{4}G^{\mu}_{\lambda}G^{\lambda}_{\nu}+\frac{1}{12}G^{\lambda}_{\lambda}G^{\mu}_{\nu}+
\frac{1}{8}G^{\lambda}_{\rho}G^{\rho}_{\lambda}\delta^{\mu}_{\nu}-\frac{1}{24}(G^{\lambda}_{\lambda})^{2}
\delta^{\mu}_{\nu}. \nn\label{pi} \ee
 It is obvious that by taking the trace of
Eqs.~(\ref{torii}), the electric components of the Weyl tensor,
which from the braneworld viewpoint are not intrinsic quantities,
disappear, and the resulting equation is \be
R^{\mu}_{\nu}R^{\nu}_{\mu}-\frac{1}{3}R^{2}+\frac{4}{r_{c}^{2}}\Big(1+\frac{\lambda}{6}r_{c}\kappa_{5}^{2}\Big)
R-\frac{8}{r_{c}^{2}}\Big(\Lambda_{5}+\frac{\lambda^{2}}{6}\kappa_{5}^{4}\Big)=0\,.
\label{traceing} \ee We can now check by substituting the
candidate solution (\ref{tre}) into Eq.~(\ref{traceing}) that this
is not satisfied, which means that the interior solution cannot
match to any static exterior.
\par
Note that, while the Friedmann equation (\ref{igr}) was initially derived in \cite{deffayet} under the
assumption of a particular bulk ansatz inducing, of course, the metric (\ref{intermetric}) on the brane,
however, the derivation of this Friedmann equation was also given in \cite{mmt} in a bulk-independent way, thus,
showing the full generality of the cosmology (\ref{igr}).

\subsection{Gauss-Bonnet term}
The cosmology of the braneworld Gauss-Bonnet model ($r_{c}\to 0$)
is given by the Friedmann equation \cite{charmousis,dl}
 \bea
\Big(\frac{\dot{a}}{a}\Big)^{2}= -\frac{k}{a^{2} }+
\frac{1}{8\alpha}\left(-2+\frac{64I ^{2}}{J}+J\right)\,,
\label{friedgb}
 \eea
where the dimensionless quantities $I , J$ are given by
 \bea
&&\!I=\frac{1}{8} \left[1+\frac{4}{3}\alpha\Lambda_{5}+
 \frac{8\alpha \mathcal{C}}{a ^{4}}\right]^{1/2}\!,
 \label{phi1}\\
&&\!\! J=\!\!
 \left[
\frac{\kappa^{2}_{5}\sqrt{\alpha}}{\sqrt{2}} (\rho +\lambda)\! +\!
\sqrt{{\kappa^{4}_{5}\alpha \over 2} (\rho + \lambda)^{2}
 \!+\!(8I )^{3} } \right]^{\!2/3}\!\!.\label{phi11}
 \eea
 Similarly to the induced gravity case, we obtain
 \be
A(r)=1+\frac{r^{2}}{8\alpha}\Big[2-\frac{64I(r)^{2}}{J(r)}-J(r)\Big]\,,
 \label{je}
 \ee
 where
 \bea
&&\!\!\!\!\!\!\!\!\!\!\!\!\!I(r)=\frac{1}{8}
\left[1+\frac{4}{3}\alpha\Lambda_{5}+
 \frac{8\alpha q}{r ^{4}}\right]^{1/2}\!,
 \label{phi1}\\
&&\!\! \!\!\!\!\!\!\!\!\!\!\!J(r)\!=\!\!
 \left[
\frac{\kappa^{2}_{5}\sqrt{\alpha}}{\sqrt{2}} \Big(\frac{m}{r^3}
\!+\!\lambda\Big)\! \!+\! \sqrt{{\kappa^{4}_{5}\alpha \over 2}
\Big(\frac{m}{r^3} \!+\! \lambda\Big)^{2}
 \!\!+\!(8I(r) )^{3} } \right]^{\!\frac{2}{3}}\!.\label{phi11}
 \eea
 \par
The generic effective braneworld equations for the Gauss-Bonnet
model have been derived in \cite{mato} in the form  \bea
\frac{3}{2}(M_{\mu\nu}+\mathcal{E}_{\mu\nu})-\frac{1}{4}Mh_{\mu\nu}+\alpha[H^{(1)}_{\mu\nu}
+H^{(2)}_{\mu\nu}+H^{(3)}_{\mu\nu}]\nn\\
+\frac{\alpha(10\Lambda_{5}-\alpha \textrm{I})}{2(3+\alpha
M)}\Big(M_{\mu\nu}-\frac{1}{4}Mh_{\mu\nu}\Big)=\frac{\Lambda_{5}}{4}h_{\mu\nu}\,,
\label{eqgb} \eea where \bea
H^{(1)}_{\mu\nu}\!\!\!&=&\!\!\!2M_{\mu\alpha\beta\gamma}M_{\nu}^{\,\,\,\alpha\beta\gamma}\!\!-\!
6M^{\rho\sigma}\!M_{\mu\rho\nu\sigma}\!+\!4M\!M_{\mu\nu}\!-\!8M_{\mu\rho}\!M_{\nu}^{\,\,\,\rho}\nn\\&&
\!\!\!\!-\frac{1}{8}h_{\mu\nu}(7M^2\!-\!24M_{\alpha\beta}M^{\alpha\beta}\!+\!3
M_{\alpha\beta\gamma\delta}M^{\alpha\beta\gamma\delta})\,,
\label{eta1} \eea \bea
\!\!\!\!\!\!\!\!\!\!\!\!\!\!\!\!\!\!\!\!\!\!\!\!\!\!\!\!\!\!\!\!\!\!\!
H^{(2)}_{\mu\nu}\!\!&=&\!\!-6(M_{\mu\rho}\mathcal{E}^{\rho}_{\nu}+M_{\nu\rho}\mathcal{E}^{\rho}_{\mu}
+M_{\mu\rho\nu\sigma}\mathcal{E}^{\rho\sigma})\nn\\
&&\!\!+\frac{9}{2}h_{\mu\nu}M_{\rho\sigma}\mathcal{E}^{\rho\sigma}+3M\mathcal{E}_{\mu\nu}\,,
\label{eta2} \eea \bea H^{(3)}_{\mu\nu}\!\!\!&=&\!\!\!-4N_{\mu}
N_{\nu}\!+\!4N^{\rho}(N_{\rho\mu\nu}\!+\!N_{\rho\nu\mu})\!+\!2N_{\rho\sigma\mu}N^{\rho\sigma}_{\,\,\,\,\,\,\,\nu}\nn\\
&&\!\!\!+4N_{\mu\rho\sigma}N_{\nu}^{\,\,\rho\sigma}\!\!+\!3h_{\mu\nu}\!\Big(\!N_{\alpha}N^{\alpha}\!\!
-\!\frac{1}{2}N_{\alpha\beta\gamma}N^{\alpha\beta\gamma}\!\Big)\!,\label{eta3}
\eea \bea
\textrm{I}\!\!&=&\!\!M^2-8M_{\alpha\beta}M^{\alpha\beta}+M_{\alpha\beta\gamma\delta}M^{\alpha\beta\gamma\delta}-
8N_{\rho}N^{\rho}\nn\\
&&+4N_{\rho\sigma\kappa}N^{\rho\sigma\kappa}-12M_{\rho\sigma}\mathcal{E}^{\rho\sigma}\,.
\label{yiota} \eea
 These equations contain the quantities
$M_{\alpha\beta\gamma\delta}$, $N_{\mu\nu\rho}$, which are
expressed in terms of the induced metric $h_{\mu\nu}$ and the
extrinsic curvature $K_{\mu\nu}$ as follows \bea
M_{\alpha\beta\gamma\delta}=R_{\alpha\beta\gamma\delta}-K_{\alpha\gamma}K_{\beta\delta}+
K_{\alpha\delta}K_{\beta\gamma} \label{M}\\
M_{\alpha\beta}=h^{\gamma\delta}M_{\alpha\gamma\beta\delta}\,\,\,\,,\,\,\,\,
M=h^{\alpha\beta}M_{\alpha\beta}\\
N_{\mu\nu\rho}=K_{\nu\rho;\,\mu}-K_{\mu\rho;\,\nu}\,\,\,\,,\,\,\,\,
N_{\mu}=h^{\rho\sigma}N_{\rho\mu\sigma}\label{N}\,, \eea where
$K_{\mu\nu}$ satisfies the matching conditions \cite{davis} (for
the vacuum case) \bea
K^{\mu}_{\nu}\!\!&+&\!\!\frac{2\alpha}{3}[9\textrm{J}^{\mu}_{\nu}-2\textrm{J}\delta^{\mu}_{\nu}
-2(3P^{\mu\rho}_{\,\,\,\,\,\,\nu\sigma}+\delta^{\mu}_{\nu}G^{\rho}_{\sigma})
K^{\sigma}_{\rho}]\nn\\\!&=&\!-\frac{\kappa_{5}^{2}\lambda}{6}\delta^{\mu}_{\nu}\,,
 \label{matchgb} \eea
with \bea
3\textrm{J}^{\mu}_{\nu}\!&=&\!2KK^{\mu}_{\rho}K^{\rho}_{\nu}+K^{\rho}_{\sigma}K^{\sigma}_{\rho}K^{\mu}_{\nu}
-2K^{\mu}_{\rho}K^{\rho}_{\sigma}K^{\sigma}_{\nu}-K^2
K^{\mu}_{\nu}
\nn\\
P^{\mu\rho}_{\,\,\,\,\,\,\nu\sigma}\!&=&\!R^{\mu\rho}_{\,\,\,\,\,\,\nu\sigma}
+2\delta^{\mu}_{[\sigma}R_{\nu]}^{\rho}
+2\delta^{\rho}_{[\nu}R_{\sigma]}^{\mu}
+R\delta^{\mu}_{[\nu}\delta_{\sigma]}^{\rho}\,,\,\,\,\,\,\,\,\,\nn
\eea ($\textrm{J}=\textrm{J}^{\mu}_{\mu}$). It is remarkable that
despite the complexity of the above equations, by taking the trace
of Eq.~(\ref{eqgb}) the quantities $\mathcal{E}_{\mu\nu}$
disappear (as well as the quantities $N_{\mu\nu\rho}$ containing
covariant derivatives of $K_{\mu\nu}$ with respect to
$h_{\mu\nu}$), and the resulting equation is purely
four-dimensional \be M+\alpha(M^2-4M_{\alpha\beta}M^{\alpha\beta}+
M_{\alpha\beta\gamma\delta}M^{\alpha\beta\gamma\delta})=2\Lambda_{5}\,.
\label{tro} \ee After solving the cubic system (\ref{matchgb}) for
$K^{\mu}_{\nu}$, one has Eq.~(\ref{tro}) constructed solely out of
the induced metric $h_{\mu\nu}$.
\par
In order to do so, it is convenient to write the metric
(\ref{static}) in a form where its angular part appears as a
two-dimensional conformally Euclidean space \be
ds^{2}\!=\!-A(r)dt^{2}+\frac{dr^{2}}{A(r)}+r^{2}\,f(x_{1},x_{2})\,(dx_{1}^{2}\!+\!
dx_{2}^{2})\,, \label{statici} \ee where
$f(x_{1},x_{2})=[1+(x_{1}^{2}+x_{2}^{2})/4]^{-2}$, and
$x_{1}=2\tan(\theta/2)\sin\phi$, $x_{2}=2\tan(\theta/2)\cos\phi$.
Due to the symmetry of the metric (\ref{statici}), the components
of $K^{\mu}_{\nu}$ in the coordinates $(t,r,x_{1},x_{2})$ take the
form \be K^{\mu}_{\nu}=diag(K_{1}, K_{2}, K_{3}, K_{3})\,.
\label{compo} \ee Subtracting, now, the $tt$-equation from the
$rr$-equation of the system (\ref{matchgb}), we obtain the
separable equation \be [4\alpha r^{2} K_{3}^{2}+4\alpha (A-1)
-r^{2}]\,(K_{2}-K_{1})=0\,, \label{bu} \ee with solutions \be
K_{3}=\pm\frac{1}{\sqrt{8\alpha}}\Big[\frac{64I(r)^{2}}{J(r)}+J(r)\Big]^{1/2}
\,, \label{ena} \ee or \be K_{2}=K_{1} \,. \label{dio} \ee For the
solution (\ref{ena}), plugging back into the system
(\ref{matchgb}), we obtain a system of two equations for
$K_{1},K_{2}$, from where the following equation arises \be
2\alpha A'-r\mp\sqrt{\alpha}\kappa_{5}^{2}\lambda
r^2/2\sqrt{r^2+4\alpha(1-A)}=0\,, \label{inco} \ee (prime means
differentiation with respect to $r$). This equation is easily seen
to be inconsistent with $A(r)$ given by Eq.~(\ref{je}). For the
solution (\ref{dio}), plugging back into the system
(\ref{matchgb}), the situation is more complicated, and a
polynomial equation of fifth degree on $K_{1}$ arises: \bea
&&K_{3}=\frac{4(1-2\alpha r^{-1} A')K_{1}+
\kappa_{5}^{2}\lambda }{8\alpha  K_{1}^{2}+2(2\alpha A''-1)}\,,\label{lost}\\
&&
\!\!\!\!\!\!a_{5}K_{1}^{5}+a_{4}K_{1}^{4}+a_{3}K_{1}^{3}+a_{2}K_{1}^2+a_{1}K_{1}+a_{0}
=0, \label{ga}\eea where  \bea
a_{5}\!\!&=&\!\!2\alpha[r^2+4\alpha(1-A)]\,\,\,\,,\,\,\,\,a_{4}=\alpha\kappa_{5}^2
\lambda r^2\nn\\
a_{3}\!\!&=&\!\!(2\alpha A''-1)[r^2+4\alpha(1-A)]\nn\\
a_{2}\!\!&=&\!\!\kappa_{5}^2 \lambda r [\alpha A'+r(\alpha A''-1)]\nn\\
2a_{1}\!\!&=&\!\!1-(2\alpha A''-1)[4rA'+A(2\alpha A''-1)-4\alpha
A'^2]\nn\\
&&\!\!+ A''[r^2-4\alpha+\alpha(r^2+4\alpha)A'']\!-\!(3+\alpha\kappa_{5}^4\lambda^2)r^2/4\alpha\nn\\
a_{0}\!\!&=&\!\!\kappa_{5}^2\lambda r(2\alpha A''-1)(2\alpha r
A''+r-4\alpha A')/16\alpha.\label{def} \eea
\par
Supposed that equation (\ref{tro}) is valid, for the metric
(\ref{statici}), and after substituting the values of $K_{2},
K_{3}$ from Eqs.~(\ref{dio}), (\ref{lost}), it becomes a
polynomial equation of sixth degree on $K_{1}$ \bea &&
\!\!\!\!\!\!b_{6}K_{1}^{6}+b_{4}K_{1}^{4}+b_{3}K_{1}^{3}+b_{2}K_{1}^2+b_{0}
=0, \label{gaa}\eea where \bea
\!\!b_{6}\!\!&=&\!\!32\alpha^2[r^2+4\alpha(1-A)]\nn\\
\!\!b_{4}\!\!&=&\!\!16\alpha[(2\alpha\Lambda_{5}-3)r^2-6\alpha-12\alpha
A'(\alpha A'-r)\nn\\
&&\,\,\,\,\,\,\,\,\,\,\,+3\alpha (r^2+4\alpha)A''-6\alpha A(2\alpha A''-1)]\nn\\
\!\!b_{3}\!\!&=&\!\!32\alpha\kappa_{5}^2 \lambda r (2\alpha A'-r)\nn\\
\!\!\frac{b_{2}}{2}\!\!\!&=&\!\!12\alpha\!-\!(3\!+\!3\alpha
\kappa_{5}^4 \lambda^2\!+\!8\alpha \Lambda_{5})r^2
\!-\!12\alpha A(2\alpha A''-1)^2\nn\\
&&\,\,\,\,\,\,\,\,\,+ 4\alpha^2 A''[4(\Lambda_{5}r^2-3)+3(r^2+4\alpha)A'']\nn\\
\!\!b_{0}\!\!\!&=&\!\!\!(2\alpha
A''\!-\!1)\{2\alpha(r^2\!+\!4\alpha)A''^2\!-\!A''[8\alpha\!+\!(1\!-\!4\alpha
\Lambda_{5})r^2]\nn\\
&&\,\,\,\,\,\,\,\,\,\,\,\,\,\,\,\,\,\,\,\,-2(2\alpha
A''\!-\!1)[A(2\alpha
A''\!-\!1)\!+\!2A'(\alpha A'\!-\!r)]\nn\\
&&\,\,\,\,\,\,\,\,\,\,\,\,\,\,\,\,\,\,\,\,+2\!-\!
(2\Lambda_{5}+\kappa_{5}^4\lambda^2/2) r^2 \}.\label{def} \eea
 Equations (\ref{ga}),
(\ref{gaa}) have to be satisfied simultaneously. After some
algebraic manipulations, this system of equations is
written equivalently as the following system \bea &&F_{2}K_{1}^2+F_{1}K_{1}+1=0\,,\label{equiv1}\\
&& K_{1}=\frac{C_2 - (C_1 - F_1)F_1- F_2}{(C_1 - F_1)
F_2-C_3}\,,\label{equiv2} \eea where
  \bea
&&\!\!\!\!\!\!\!\!(C_1,C_2,C_3)\!=\!\!\Big(\!B_2(B_1\!-\!p_1\!)
\!+\!p_3\!-\!B_{3},B_3(B_1\!-\!p_1\!)\!+\!p_4\!-\!B_4,\nn\\
&&\,\,\,\,\,\,\,\,\,\,\,\,\,\,\,\,\,\,\,\,\,\,\,\,\,\,\,\,\,\,B_4(B_1\!-\!p_1)+p_5\!\Big)/(B_1(B_1\!-\!p_1)
\!+\!p_2\!-\!B_{2}), \nn\\
&&\!\!\!\!\!\!\!(F_1\,,\,F_2)\!=\!\Big(\!(C_1
C_2\!-\!C_3)(B_1\!-\!p_1)
\!-\!C_2 (B_2\!-\!p_2)\!+\!B_4\!-\!p_4,\nn\\
&&\,\,\,\,\,\,\,\,\,\,\,\,\,\,\,\,\,\,\,\,\,\,\,\,\,\,\,\,C_1
C_3(B_1\!-\!p_1)
\!-\!C_3 (B_2\!-\!p_2)\!-\!p_5\!\Big)/\nn\\
&&\,\,\,\,\,\,\,\,\,\,\,\,\,\,\,\,\,\,\,\,\,\,\,\,\,\,\,\Big(\!(C_1^2\!-\!C_2)(B_1\!-\!p_1)
\!-\!C_1 (B_2\!-\!p_2)\!+\!B_3\!-\!p_3\!\Big)\!,\nn\\
 &&\!\!\!\!\!\!\!(B_1\,,\,B_2\,,\,B_3\,,\,B_4)\!=\!(p_1 p_2\!-\!p_3\!+\!q_3\,,
 \,p_1 p_3\!-\!p_4\!+\!q_4\,,\nn\\
&&\,\,\,\,\,\,\,\,\,\,\,\,\,\,\,\,\,\,\,\,\,\,\,\,\,\,\,\,\,\,\,\,
\,\,\,\,\,\,\,\,\,\,\,\,\,\,\,\,\,p_1 p_4\!-\!p_5 \,,\,p_1
p_5\!+\!q_6)/(p_1^2\!-\!p_2\!+\!q_2),\nn \eea and the various
$p_{i}$, $q_{i}$ are related to $a_{i}$, $b_{i}$ as \be
p_{i}=a_{i}/a_{0}\,\,\,\,\,\,,\,\,\,\,\,\,q_{i}=b_{i}/b_{0}\,.
\label{final}\ee It is now straightforward to check that the value
$K_{1}$ of Eq.~(\ref{equiv2}) does not satisfy Eq.~(\ref{equiv1}),
which means that the no-go theorem of the Gauss-Bonnet braneworld
has been proved.

\subsection{Gauss-Bonnet and Induced gravity}
The cosmology of the combined Gauss-Bonnet and induced gravity
braneworld is given by the Friedmann equation \cite{kmp} \bea &&
\Big(\frac{\dot{a}}{a}\Big)^{2}=-\frac{k}{a ^{2}} +\frac{4-3\beta
}{12\beta\alpha } \nn\\&&~~~{}-\frac{2}{3\beta\alpha } \sqrt{P
^{2}-6 Q }\,\cos\left(\Theta\pm \frac{\pi}{3}\right)\,,
\label{mama}
 \eea
where the dimensionless quantities $\beta, P, Q, \Theta$ are given
by
 \bea
\beta& = & { 256\alpha \over 9r_{c}^2}\,,\label{beta}\\
 P & = & 1+3\beta
I\,,\label{parameterP}
\\  Q
&=&\beta\left[\frac{1}{4} +I+\frac{\kappa_{4}^{2}\alpha}{3}
(\rho+\lambda)\right],\label{parameterQ}\\ \Theta( P ,Q
)&=&\frac{1}{3}\arccos\left[\frac{2P^{3}+27Q^{2}-18PQ}
{2(P^{2}-6Q)^{3/2}}\right] \!. \label{omega}
 \eea
 The $\pm$ sign in Eq.~(\ref{mama}) is the same as that in
 Eq.~(\ref{igr}). The region in ($ P ,Q $)-space for which Eq.~(\ref{mama}) is
defined, is
 \bea
&& 1\leq P <{4 \over 3}\,, \label{ineq1}\\ && 2[\,9P -8-(4-3P
)^{3/2}\,]\leq 27 Q \nn
\\&&~~~~{} \leq 3P [\,3-\sqrt{3(3-2P )}\,]\,.\label{ineq2}
 \eea
 From the above equations, we can write the candidate black hole
 metric as
 \bea
&& A(r)=1-\frac{4-3\beta }{12\beta\alpha }r^2
\nn\\&&~~~{}+\frac{2r^2}{3\beta\alpha } \sqrt{P(r) ^{2}-6 Q(r)
}\,\cos\left(\Theta(r)\pm \frac{\pi}{3}\right)\,,
 \label{bh}
 \eea
where \bea \!\!\!\!\!\!\!\!P(r) \!\!& = &\!\! 1+3\beta
I(r)\,,\label{parameterP}
\\  \!\!\!\!\!\!\!\!Q(r)
\!\!&=&\!\!\beta\left[\frac{1}{4}
+I(r)+\frac{\kappa_{4}^{2}\alpha}{3}
\Big(\frac{m}{r^3}+\lambda\Big)\right],\label{parameterQ}\\
\!\!\!\!\!\!\!\!\Theta(r)\!\!&=&\!\!\frac{1}{3}\arccos\!\left[\!\frac{2P(r)^{3}+27Q(r)^{2}-18P(r)Q(r)}
{2(P(r)^{2}-6Q(r))^{3/2}}\!\right] \!. \label{omega}
 \eea
Similarly to the induced gravity case, in the pure Gauss-Bonnet model, the Friedmann equation (\ref{friedgb})
was obtained in \cite{mato} in a bulk-independent way, showing, thus, that this is the most general
Robertson-Walker braneworld cosmology of the model. For the combined Gauss-Bonnet and induced gravity braneworld
scenario, its Friedmann equation (\ref{mama}) has not been derived yet in a bulk-independent way, but we have no
reason to expect any discrepancy in this case too.
\par
In the present case, Eqs.~(\ref{eqgb})-(\ref{N}) -- as well as
Eq.~(\ref{tro}) -- of the pure Gauss-Bonnet case remain unchanged,
since they are described by bulk information, while the matching
condition (\ref{matchgb}) is now modified by setting its right
hand side equal to \be -\frac{\kappa_{5}^2}{6}\Big[\lambda
\delta^{\mu}_{\nu}-\frac{3r_{c}}{\kappa_{5}^2}\Big(R^{\mu}_{\nu}-\frac{R}{6}\delta^{\mu}_{\nu}\Big)\Big]\,.
\ee Following the same steps as before, we find the same separable
equation (\ref{bu}), with solutions now \bea
K_{3}\!=\!\pm\frac{1}{\sqrt{3\beta\alpha}}\Big[1\!-\!2\sqrt{P(r)^{2}\!-\!6
Q(r) }\cos\!\Big(\!\Theta(r)\!\pm\!
\frac{\pi}{3}\!\Big)\Big]^{\frac{1}{2}}\!, \label{enaco} \eea or
\be K_{2}=K_{1} \,. \label{dioco} \ee Solution (\ref{enaco}) leads
to the equation \bea &&\!\!\!\!\!\!\!2\alpha
A'\!-\!r\!\mp\!\sqrt{\alpha}[\kappa_{5}^2\lambda
r^2\!-\!r_{c}(1\!-\!A\!-\!r
A')]/2\sqrt{r^2\!+\!4\alpha(1\!-\!A)}\nn\\
&&\,\,\,\,\,\,\,\,\,\,\,\,\,\,\,\,\,\,\,\,\,\,\,\,\,\,\,\,\,\,
\,\,\,\,\,\,\,\,\,\,\,\,\,\,\,\,\,\,\,\,\,\,\,\,\,\,\,\,\,\,
\,\,\,\,\,\,\,\,\,\,\,\,\,\,\,\,\,\,\,\,\,\,\,\,\,\,\,\,\,\,
\,\,\,\,\,\,\,\,\,\,\,\,\,\,\,\,\,\,\,\,=0\,, \label{incoco} \eea
which is inconsistent with Eq.~(\ref{bh}). Solution (\ref{dioco})
leads to two equations of fifth and sixth degree on $K_{1}$ of the
form (\ref{ga}) and (\ref{gaa}) respectively, with the
corresponding coefficients defined now as follows (denoted with
primes) \bea
a'_{5}\!\!&=&\!\!a_{5}\,\,\,\,,\,\,\,\,a'_{4}=a_{4}-\alpha
r_{c}(1-A-r A')\,\,\,\,,\,\,\,\,
a'_{3}=a_{3}\nn\\
a'_{2}\!\!&=&\!\!a_{2}+r_{c}[1+2\alpha A'^2+A(2\alpha A''-1)+r A' (3\alpha A''-2)\nn\\
&&\,\,\,\,\,\,\,\,\,\,\,\,\,\,\,\,\,\,\,\,\,\,\,\,\,
\,\,\,\,\,\,\,\,\,\,\,\,\,\,\,\,\,\,\,\,\,\,\,\,
\,\,\,\,\,\,\,\,\,\,\,\,\,\,\,\,\,\,\,\,\,\,\,\,\,\,\,\,\,\,-A''(2\alpha+r^2/2)]/2\nn\\
a'_{1}\!\!&=&\!\!a_{1}-r_{c}(2 A'+r A'')[4\kappa_{5}^2\lambda r+r_{c}(2A'+r A'')]/32\nn\\
a'_{0}\!\!&=&\!\!a_{0}\nn\\
b'_{6}\!\!&=&\!\!b_{6}\,\,\,\,,\,\,\,\,b'_{4}=b_{4}\nn\\
b'_{3}\!\!&=&\!\!b_{3}-16\alpha r_{c}(r-2\alpha A')(2A'+r A'')\nn\\
b'_{2}\!\!&=&\!\!b_{2}-3\alpha r_{c}(2 A'+r A'')[4\kappa_{5}^2\lambda r+r_{c}(2A'+r A'')]/2\nn\\
b'_{0}\!\!&=&\!\!b_{0}.\label{def} \eea Equations
(\ref{equiv1})-(\ref{final}) remain the same for the above primed
quantities $a'_{i}, b'_{i}$, and thus, their incompatibility can
be easily checked, proving the no-go theorem of the combined
Gauss-Bonnet and induced gravity braneworld.
\par
Our analysis of all the above considered models is based on 4-dimensional solutions or
4-dimensional effective braneworld equations, and we have not studied the bulk extension of the
considered braneworld regions (static exterior and Robertson-Walker interior). We know however,
that for a given continuous boundary metric and extrinsic curvature, the propagation of the field
equations in five dimensions is a well defined initial value problem, solvable in principle.
\par
In conclusion, we have studied the Oppenheimer-Snyder-like
collapse on braneworld models with curvature corrections. In all
cases considered, using the four-dimensional effective equations,
and without making assumptions about the bulk, we have found that
the exterior vacuum spacetime on the brane is non-static. We have
not found the exterior metric, thus, we are not in position to
know if the gravitational collapse on the brane leaves at late
times a signature in the exterior, or if, on the contrary, the
non-static exterior is transient, tending to a static geometry.

\[ \]
{\bf Acknowlegements}

We wish to thank R. Maartens for useful discussions. G.K.
acknowledges partial support from FONDECYT grant 3020031 and from
Empresas CMPC. Centro de Estudios Cient\'{\i}ficos is a Millennium
Science Institute and is funded in part by grants from Fundacion
Andes and the Tinker Foundation. E.P. is partially supported by
the NTUA research program ``Thalis".

\end{document}